\documentclass[structabstract]{aa}  
\usepackage{natbib}
\bibpunct{(}{)}{;}{a}{}{,} 
\usepackage[german,english]{babel}
\usepackage{graphicx}
\usepackage{epsfig}
\usepackage{xspace}
\usepackage{epsf}
\usepackage{epstopdf}
\usepackage{amsmath}
\usepackage{txfonts}
\begin{document}
   \title{The bright-end of the luminosity function at $z\sim$9
   \thanks{Based on observations collected at The European Southern
     Observatory, Paranal, Chile, as part of the ESO 082.A-0163 and 087.A-0118}}


 \author{N.~Laporte\inst{1,2}
          \and
R.~Pell\'o\inst{1,2}\and
M.~Hayes\inst{1,2} \and
D.~Schaerer\inst{5,2} \and
F.~Boone\inst{1,2} \and
J.~Richard\inst{4}  \and
J.F.~Le Borgne\inst{1,2} \and
J.P.~Kneib\inst{3} \and
F. Combes\inst{6}
         }


\institute{
Universit\'e de Toulouse; UPS-OMP; IRAP; Toulouse, France\\
\and
CNRS; IRAP; 14, avenue Edouard Belin, F-31400 Toulouse, France\\
\email{nicolas.laporte@irap.omp.eu,roser@irap.omp.eu,matthew.hayes@irap.omp.eu,frederic.boone@irap.omp.eu}
\and
Laboratoire d'Astrophysique de Marseille, CNRS - Universit\'e Aix-Marseille,
38 rue Fr\'ed\'eric Joliot-Curie, 13388 Marseille Cedex 13, France \\
\email{jean-paul.kneib@oamp.fr}
\and
Centre de Recherche Astrophysique de Lyon, University Lyon 1, 9 Avenue
Charles Andr\'e, 69561 Saint Genis Laval, France \\
\email{johan.richard@univ-lyon1.fr}
\and
Observatoire de Gen\`eve, Universit\'e de Gen\`eve, 51 Ch. des Maillettes, 1290 Versoix, Switzerland\\
\email{Daniel.Schaerer@unige.ch}
\and
LERMA, Observatoire de Paris and CNRS, 61 Avenue de l'Observatoire, 75014 Paris, France\\
\email{Francoise.Combes@obspm.fr}
	  }

   \date{Received ; accepted }

   \abstract
   {We report new constraints on the galaxy luminosity function at $z\sim$9 based on observations carried out with ESO/VLT FORS2, HAWK-I and X-Shooter around the lensing cluster A2667, as part of our project aimed at selecting $z\sim$ 7-10 candidates accessible to spectroscopy. Only one $J-$ dropout source was selected in this field fulfilling the color and magnitude criteria. This source was recently confirmed as a mid-$z$ interloper based on X-Shooter spectroscopy.}
  {The depth and the area covered by our survey are well suited to set strong constraints on the bright-end of the galaxy luminosity function and hence on the star formation history at very high redshift.}{The non-detection of reliable $J-$ dropout sources over the $\sim$36arcmin$^2$ field of view towards A2667 was used to carefully determine the lens-corrected effective volume and the corresponding upper-limit on the density of sources.}
{The strongest limit is obtained for $\Phi$(M$_{1500}$=-21.4$\pm$0.50)$<$6.70$\times$10$^{-6}$Mpc$^{-3}$mag$^{-1}$ at $z\sim$9. A maximum-likelihood fit of the luminosity function using all available data points including the present new result yields M$^{\star}>$-19.7 with fixed $\alpha$=-1.74 and $\Phi^{\star}$=1.10$\times$10$^{-3}$Mpc$^{-3}$. The corresponding star formation rate density should be $\rho_{SFR}<$5.97$\times$10$^{-3}$M$_{\odot}$ yr$^{-1}$ Mpc$^{3}$ at $z\sim$9. These results are in good agreement with the most recent estimates already published in this range of redshift and for this luminosity domain.}
{This new result confirms the decrease in the density of luminous galaxies at very high-redshift, hence providing strong constraints for the design of future surveys aiming to explore the very high-redshift Universe.}



   \maketitle
%

\section{Introduction}

   Understanding the formation and evolution of the first galaxies is one of the major topics of modern astronomy. Since a dozen years, many projects and instruments have been developped to push the boundaries of the observed Universe, such as the Hubble Ultra Deep Field \citep{2003AAS...202.1705B} for example. The role played by the first luminous objects in the Universe during the cosmic reionization is still unclear. One way to explore the first epochs is to constrain the number density of galaxies along the cosmic time, and thus establish the Star Formation History (SFH) which represents the evolution of the Star Formation Rate density (SFRd) over the cosmic time. Two complementary approaches could be adopted to further constrain the luminosity function : large blank fields to select the brightest sources, and lensing clusters used as gravitational telescopes to select the faintest ones. Our group has conducted an observing program combining these two approches aimed at selecting $z\sim$ 7-10 candidates accessible to spectroscopy, namely the WIRCAM Ultra Deep Survey (WUDS) at CFHT on the CFHTLS-D3 Groth Strip field (Pello et al., to be submitted), and a lensing survey (\citealt{Laporte2011}, hereafter L11), which is central to this letter.

   The galaxy Luminosity Function (LF) is relatively well constrained up to $z\sim$6. Beyond this limit however the contamination by low-$z$ interlopers becomes more severe because a majority of samples are only supported by photometric considerations, and therefore the results at these redshifts could be seriously affected by contamination \citep{2011A&A...534A.124B}.
The only way to get rid of this problem is to spectroscopically confirm, if not all, at least a substantial fraction of sources in a given sample, a performance beyond the capabilities of present-day facilities. Only the brightest samples, including highly magnified sources, could be presently confirmed. Fortunately, the intermediate-to-bright accessible domain (M$_{UV}\sim$-20.5 to -22) is also the most sensitive region of the LF, where most of the evolution is currently observed as a function of redshift (e.g. Bouwens et al.\ 2010). However, at $z\sim$7-8 the LF started to be set with recent observations from ground based or space observatories (such as \citealt{2012arXiv1201.0755O}, \citealt{2011MNRAS.418.2074M}, \citealt{2010A&A...524A..28C}). Some attempts have been tried up to $z\sim$10 but without spectroscopic confirmation \citep{2011Natur.469..504B}.
   
    Deep images of the lensing cluster A2667 were obtained with ESO/VLT FORS 2 and HAWK-I in $I$, $z$, $Y$, $J$, $H$ and $Ks$ bands. All details regarding these observations and data reduction can be found in L11. IRAC data between 3.6 and 8$\mu $m, and MIPS 24$\mu $m were also included in the analysis when available (Spitzer Space Telescope, \citealt{2004ApJS..154....1W}, \citealt{2004ApJS..154...10F}, \citealt{2004ApJS..154...25R}). Based on these data, 10 photometric candidates at $z\sim$7-10 were selected based on the usual dropout technique. Among them, only one candidate, J1, is consistent with the z$\sim$9 $J-$drop selection criteria, with observed m$_H$=25.21$\pm$0.08 and a modest magnification factor $\mu$=1.3. Despite an excellent fit of the photometric SED at high-$z$, the reliability of this source was already considered as dubious by L11 given the detection of a faint compact counterpart on the HST F850LP/ACS image, with $z_{850}=$27.39$\pm$0.18. J1 was recently confirmed as a $z$=2.082 interloper based on the detection of 5 emission lines ( $[$OIII$]\lambda$5007,4959, H$\alpha$, H$\beta$ and Ly$\alpha$) with ESO/VLT X-Shooter spectroscopy (see \citealt{Hayes2012}, submitted, for more details).  

   \begin{figure*}[!]
   \centering
   \includegraphics[width=0.55\textwidth,angle=-90]{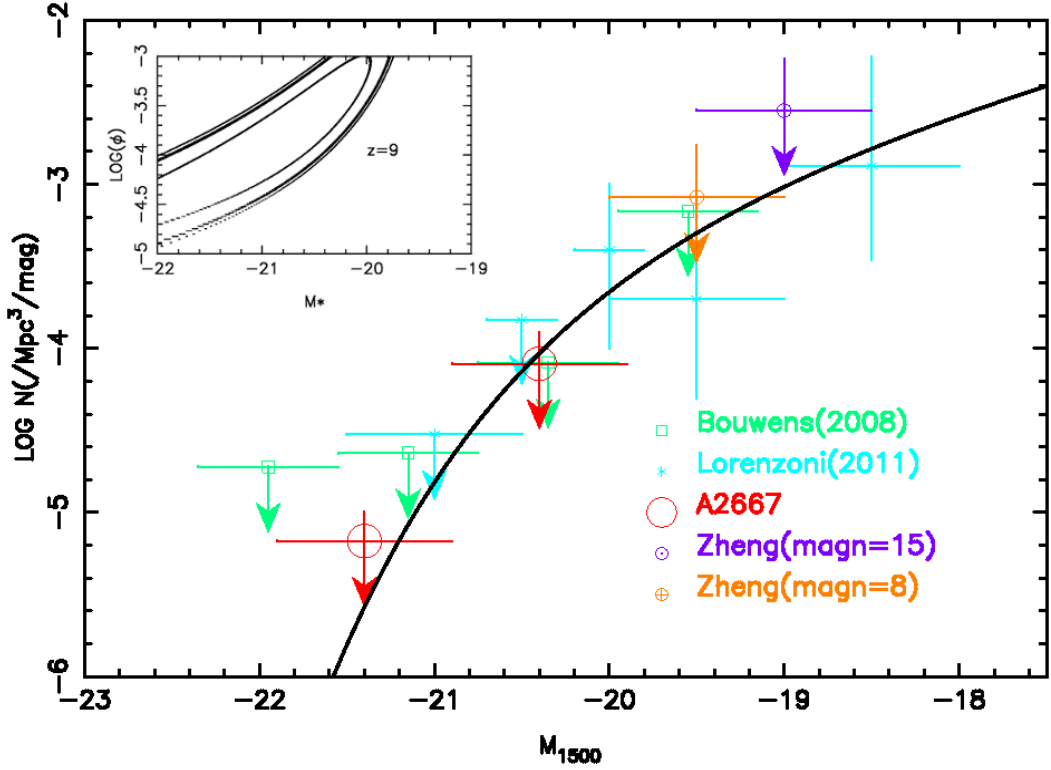}
      \caption{Luminosity function at $z\sim$9 showing the upper limits resulting from the absence of $J$-dropout sources in our HAWK-I survey (red dots including cosmic variance), upper limit from \citet{2008ApJ...686..230B} found in the HUDF and GOODS field of view (green dot). We also plot points obtained from the $z\sim$9 candidate of \citet{2012arXiv1204.2305Z}  considering a magnification factor of $\mu\sim$15, and the amplification factor $\mu\sim$8 obtained using the model by \citet{2009ApJ...707L.163S} for MACS1149+22(error bars take into account the cosmic variance computed from \citet{2008ApJ...676..767T}) and points from \citet{2011MNRAS.414.1455L} at $z\sim$8-9.  The last data set is only displayed for comparison purposes; it was not used for the computation of the LF at $z\sim$9 given the difference in the redshift domain. The dark line displays the Schechter function using M$^{\star}$=-19.7, $\alpha$=-1.74 and $\Phi^{\star}$=1.10$\times$10$^{-3}$Mpc$^{-3}$mag$^{-1}$. The small panel displays the likelihood contours for the 1, 2 and 3 $\sigma$ confidence regions for a fixed $\alpha$=-1.74. 
}
         \label{LFz9_plot}
   \end{figure*}
%

	In this letter we report on new constraints on the LF at $z\sim$9 based on the non-detection of realiable $J-$dropout candidates over the field of view towards the lensing cluster A2667. The largest magnification regimes cannot be addressed with a single cluster because the effective surfaces/volumes explored in this domain are too small to retrieve significant results. Therefore, this study is limited to the low to moderate magnification regime, where this pilot study provides the most interesting results on the LF as compared to previous studies, in particular blank-field surveys. In Sect.\ \ref{results} we introduce the method used to constrain the bright-end of the LF at $z\sim$9, in particular to determine the lens-corrected effective volume and the corresponding upper-limit on the density of sources. The constraints derived on the SFH at $z\sim$9 are presented in Sect.\ \ref{SFRd}. A brief discussion and 
conclusions are given in Sect.\ \ref{conclusion}. Throughout this paper, a concordance cosmology is adopted, with $\Omega_{\Lambda}=0.7$, $\Omega_{m}=0.3$ and $H_{0}=70\ km\ s^{-1}\ Mpc^{-1}$. All magnitudes are given in the AB system \citep{1983ApJ...266..713O}.


\section{The luminosity function at $z\sim$9}
\label{results}

    The non-detection of realiable $J-$ dropout candidates over the field of view toward A2667 is used in this Section to carefully determine the lens-corrected effective volume at $z\sim$9 and the corresponding upper-limits on the density of sources. In the following, we only consider the field of view corresponding to the surface covered with more than 75\% of the total exposure time in all filters. A mask was also applied to remove all the noisy regions within and around galaxies from the subsequent analysis. The region masked is only $\sim$12\% of the total area, but it reaches a maximum of $\sim$30\% in the cluster core, within the 1arcmin$^2$ centered on the cD galaxy. The final effective area is $\sim$36arcmin$^2$.

    The selection criteria adopted by L11 for $8.5\lesssim z \lesssim 9.5$ candidates require a detection above 5$\sigma$ level within a 1.3\arcsec \ diameter aperture in $H$, i.e. $H<$ 26.22, and the usual color selection as follows: $J-H>0.76$, $H-Ks<0.5$, and $J-H>1.3\times(H-Ks)+0.76$. In practice, the depth in the $J-$band limits the reliability of the $J-$dropout selection to $H<$ 26.0, corresponding to a completeness level higher than 90\%. Absolute magnitudes in the $\sim$1500\AA \ domain (M$_{1500}$) are derived from $H-$band magnitudes assuming a flat spectrum.  

    As compared to blank fields, data obtained in gravitational-lensing fields need to be corrected for both magnification and dilution effects. A magnification factor $\mu$ introduces an enhancement on the observed luminosity, depending on the redshift of the source and the location of the image on the sky, without modifying the color-selection window. Dilution reduces the effective surface covered by the survey, in such a way that a pixel affected by a magnification factor $\mu$ has an effective area reduced by the same factor on the corresponding source plane. We have used the lensing model for A2667 originally obtained by \citet{2006A&A...456..409C} to compute the magnification map at $z\sim$9 using the public software \textit{Lenstool}\footnote{{\tt http://www.oamp.fr/cosmology/lenstool}}, including the MCMC optimization method of \citet{2007NJPh....9..447J}. We explicitly compute the effective surface probed by the survey through a pixel-to-pixel integration of the magnification map, after masking the pixels lying in the object's mask, as explained above. The effective surface is 32.55 arcmin$^2$, corresponding to a covolume of 57602.5 Mpc$^3$ between $z\sim$8.5 and 9.5. 

    The effective limiting magnitude for the detection of $z\sim$9 candidates, however, depends on the magnification factor $\mu$. To take this effect into account when deriving upper-limits on the density of sources, we have considered different magnification regimes, with effective surface/covolume decreasing with increasing $\mu$. Table\ \ref{LF_points} summarizes these results and provides the non-detection constraints obtained on the LF. Areas, volumes and limiting magnitudes quoted in this table are effective lens-corrected ones. Given the low to moderate magnification regimes explored in this study, uncertainties in the effective surveyed volumes due to lensing-modeling uncertainties are considered negligible (see also Maizy et al. 2010 for details). The strongest limit is obtained for $\Phi$(M$_{1500}$=-21.4$\pm$0.50)$<$6.70$\times$10$^{-6}$Mpc$^{-3}$ at 68\% confidence level assuming Poissonian statistics. Figure\ \ref{LFz9_plot} displays the independent data points on the LF, together with previous results from the literature. We also computed points from \citet{2012arXiv1204.2305Z}, for comparison purpose, using MC approach taking into account the probability distribution in redshift \citep{2002A&A...395..443B} and considering two different values for the magnification : $\mu\sim$15, as reported by the authors, and $\mu\sim$8, as obtained from the MACS1149+22 model by \citet{2009ApJ...707L.163S}. 

\begin{table}
\caption{\label{LF_points} Upper limits for the luminosity function at $z=$8.5-9.5, for
different magnifications regimes and UV luminosities.}
\centering
\begin{tabular}{p{0.4cm}p{0.4cm}p{0.4cm}ccp{0.6cm}p{0.9cm}c}
\hline\hline
$\mu$ & $\mu$ & $\mu$ & Surface & Volume & $H<$ & M$_{1500}$ & $\Phi$$\times$10$^{-6}$ \\
mean  & min   & max   & arcmin$^2$& $\times$10$^{4}$Mpc$^{3}$ & mag &       & Mpc$^{-3}$.mag$^{-1}$ \\
\hline
1.04 & 1.02 & 1.05 & 32.55 & 5.7602 & 26.02 & -21.35 & 6.70 \\
1.12 & 1.05 & 1.2  & 27.01 & 4.7798 & 26.06 & -21.31 & 8.07 \\
1.35 & 1,2  & 1,5  &  9.47 & 1.6758 & 26.22 & -21.15 & 23.0 \\
2.40 & 1.5  & 3.3  &  2.31 & 0.4088 & 26.93 & -20.44 & 94.4 \\
\hline
\end{tabular}
\end{table}
    
    We have adopted the Schechter parametrization of the LF \citep{1976ApJ...203..297S}:
\begin{equation}
\Phi(M_{1500})=\Phi^{\star}\frac{\ln(10)}{2.5}\Big(10^{-0.4(M-M^{\star})}\Big)^{\alpha + 1} \exp{\Big(-10^{-0.4(M-M^{\star})}\Big)}
\end{equation}
to set constraints on M$^{\star}$ using a $\chi^2$ minimization, assuming a fixed $\alpha=$-1.74 \citep{2008ApJ...686..230B}. Our two independent limits in $\Phi$ at M$_{1500}=$-21.35 and -20.44 have been combined to the available upper limits from the HUDF and GOODS fields at the same redshift domain than in \citet{2008ApJ...686..230B}. As shown in Fig.\ \ref{LFz9_plot}, given the degeneracy between $\Phi^{\star}$ and M$^{\star}$, and the small number of available data points at $z\sim$9, we can only derive a limit for $\Phi^{\star}$ when assuming M$^{\star}$ or vice-versa. It has been shown that $\Phi^{\star}$ seems not evolve significatively from $z\sim$4 to 9, whereas M$^{\star}$ clearly decreases over $z\sim$4-8 (e.g. \citet{2011arXiv1109.0994B}). Assuming this no-evolution, we have used a fixed $\Phi^{\star}=$1.10$\times$10$^{-3}$Mpc$^{-3}$ and we obtain M$^{\star}>$-19.7. This result is also consistent with previous findings (\citet{2008ApJ...686..230B}).

\section{The star formation history up to $z\sim$9}
\label{SFRd}
   
    The SFH could be explored through the evolution of the SFRd over the cosmic time. Using the shape of the LF found in the previous section, we are now able to constrain the SFRd at $z\sim$9. This density is deduced from the UV density produced by Lyman Break galaxies and usually defined by : 
\begin{equation}
\rho_{UV} = \int^{\infty}_{0.05L^{\star}_{z=3}} L_{1500}\Phi(L_{1500}) dL_{1500}
\end{equation}
where $L^{\star}_{z=3}$ =1.33$\times$10$^{41}$erg s$^{-1}$ $\AA^{-1}$ and
$\Phi(L_{1500})$ is the UV LF. The lower integration bound was chosen in order to facilitate the comparison with previous studies (see Sect.\ \ref{conclusion}).
The conversion between UV density and SFRd is done using the \citet{1998ApJ...498..541K} relation given by :
\begin{equation}
\rho_{SFR}(\text{M$_{\odot}$.yr$^{-1}$.Mpc$^{-3}$})=1.40\,10^{-28}\times\rho_{UV}(\text{erg.s$^{-1}$.Mpc$^{-3}$.$\AA^{-1}$})
\end{equation}

\begin{figure}
   \centering
   \includegraphics[width=0.35\textwidth,angle=-90]{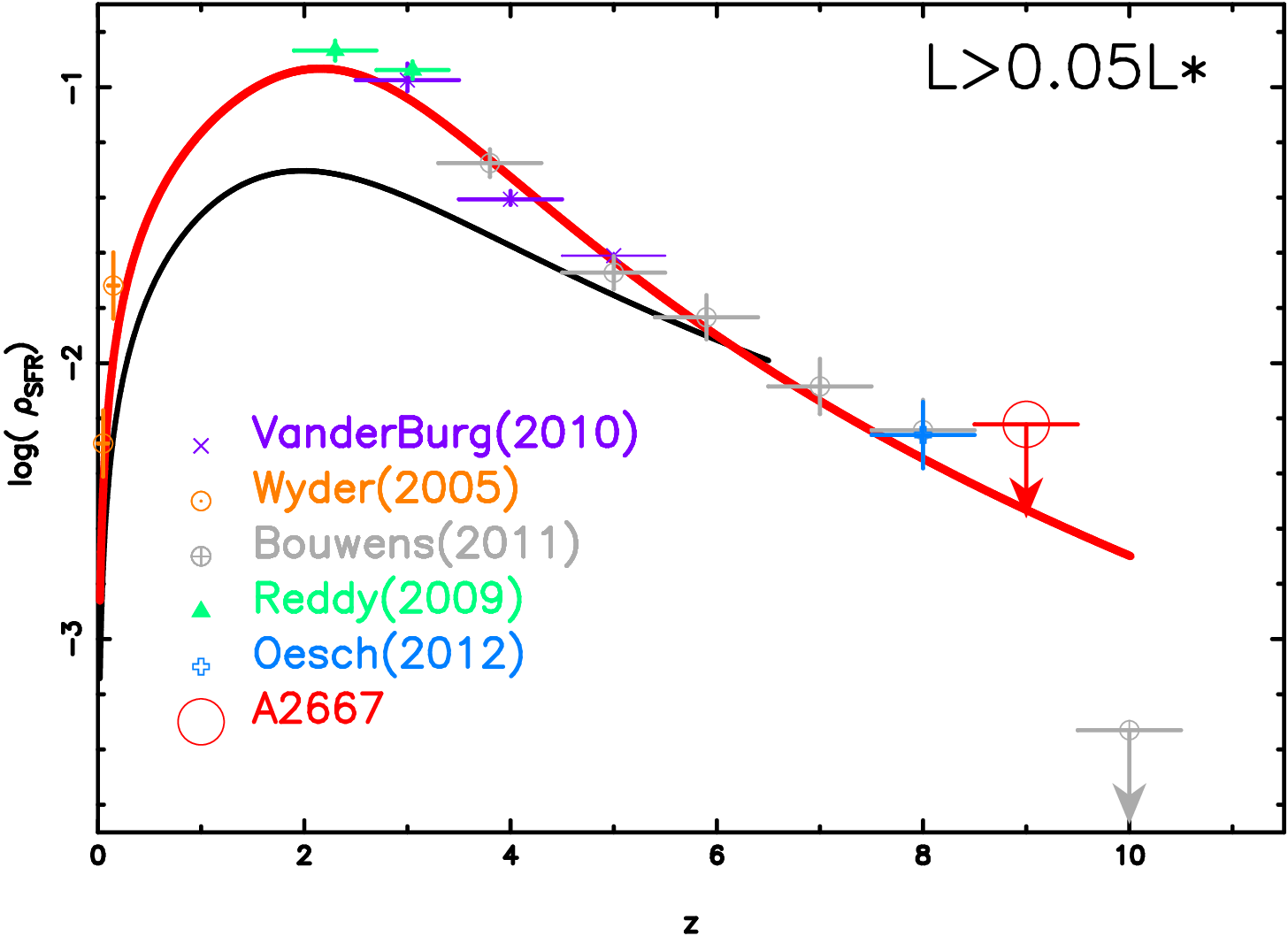}
      \caption{Star formation history including the present result at \citet{2005ApJ...619L..15W}, \citet{2009ApJ...692..778R}, \citet{2010A&A...523A..74V}, \citet{2012arXiv1201.0755O}). For consistency, all points in this figure have been corrected for dust reddening (see text). The red line shows the parametrization of the corrected SFH following \citet{2001MNRAS.326..255C} (see text) and the dark line displays the evolution of uncorrected SFRd (i.e. without dust extinction correction; the corresponding points are not shown to avoid confusion). }
         \label{SFH}
   \end{figure}

   As a result, the SFRd limit at $z\sim$9 is found, using M$^{\star}$=-19.7, to be $\rho_{SFR}<$5.97$\times$10$^{-3}$M$_{\odot}$ yr$^{-1}$ Mpc$^{3}$. Fig.~\ref{SFH} shows the evolution of the SFRd versus cosmic time including results already published over a large range in redshift. For consistency, all points in this figure up to $z\sim$7 have been corrected for dust extinction (at $z\ge$8 we considered that dust attenuation is negligible). A parameterization as a function of redshift has been derived based on the expression for dust extinction given by \citet{2005ApJ...619L..47S} as a function of the UV slope, the evolution of which as a function of redshift is taken from \citet{2005ApJ...619L..47S}  for $z\le$3.5, and extended from$z\sim$3.5 up to 7 according to \citet{2011arXiv1109.0994B}. The expression\footnote{$\rho_{\star}$ = $\frac{a+b\times z}{1+(\frac{z}{c})^d}$h \hspace{0.2cm} M$_{\odot}$.yr$^{-1}$.Mpc$^{-3}$} given by \citet{2001MNRAS.326..255C} is used to fit the SFH, in order to compare the above value of $\rho_{SFR}$ with the trend observed between $z\sim$3 and 8. The best-fit parameters of the Cole function are found to be a=0.0, b=0.10, c=2.85 and d=4.65. As seen in Fig.~\ref{SFH}, our result is compatible with the evolution observed below $z\sim$8, and it is also consistent with the SFRd limit published at $z\sim$10 by \citet{2011Natur.469..504B}.

%
    
\section{Discussion and Conclusions}
\label{conclusion}

    Over the $\sim$36arcmin$^2$ field of view of our survey around the lensing cluster A2667, only one $J-$dropout candidate was retained (L11). This source, already considered as a possible interloper by L11, has been recently confirmed at $z$=2.082 based on the detection of five emission lines with ESO/VLT X-Shooter spectroscopy (see Hayes et al., submitted, for more details). The non-detection of $J-$dropout candidates in our survey is consistent with previous non-detections reported by \citet{2008ApJ...686..230B} in the HDF, HUDF and GOODS fields (4.8, 9.1 and 9.3 arcmin$^2$ respectively), with m$_H$ (5$\sigma$, 0.6arcsec aperture) typically ranging between $\sim$27.6 to 26.6, and also complementary in terms of surface covered and effective depth.     

   Assuming that there is no reliable $z\sim$9 candidate in our survey up to m$_H$=26.0$+$2.5$\times$log($\mu$), we take benefit from the depth and large area covered to set strong constraints on the bright-end of the LF and hence on the SFH at very high redshift by a careful determination of the lens-corrected effective volume and corresponding upper-limits on the density of sources. We have considered different magnification regimes, with effective surface/covolume decreasing with increasing $\mu$. The strongest limit is obtained for $\Phi$(M$_{1500}$=-21.4$\pm$0.50)$<$6.70$\times$10$^{-6}$Mpc$^{-3}$mag$^{-1}$ at 68\% confidence level assuming Poissonian statistics. 

   We combined our two independent LF limits at M$_{1500}=$-21.35 and -20.44 with the available upper limits for the HUDF and GOODS fields, extracted from \citet{2008ApJ...686..230B}, to derive the Schechter parameters for the LF. Assuming a fixed $\alpha=$-1.74 and no-evolution in $\Phi^{\star}=$1.10$\times$10$^{-3}$Mpc$^{-3}$, we obtain a constrain on M$^{\star}>$-19.7. This result is consistent with previous findings, e.g. \citet{2008ApJ...686..230B}, \citet{2010A&A...524A..28C} (M$^{\star}$($z\sim$7)$\approx$-20.24) and \citet{2012arXiv1201.0755O} (M$^{\star}$($z\sim$8)$\approx$-19.80). The corresponding SFRd given the previous LF should be $\rho_{SFR}<$5.97$\times$10$^{-3}$M$_{\odot}$ yr$^{-1}$ Mpc$^{3}$ at $z\sim$9. These results are in good agreement with the most recent estimates already published in this range of redshift and for this luminosity domain, and confirm the decrease in the density of luminous galaxies from $z\sim$6 to 9, hence providing strong constraints for the design of future surveys aiming to explore the very high-redshift Universe.

    Several projects such as CLASH \citep{2011arXiv1106.3328P} or CANDELS \citep{2011ApJS..197...35G} are currently ongoing to constrain the evolution of galaxies beyond $z\sim$6. However, all samples highlighted by these projects need to be confirmed by deep spectroscopic observations beyond the limits of current spectrographs. Near-IR spectrographs are also needed to understand the nature of (extreme) mid-$z$ interlopers presently found in such deep surveys. The arrival of multi-object near-IR facilities such as EMIR/GTC (\citealt{2000SPIE.4008..797B}, \citealt{2011sf2a.conf..161P}) or KMOS/VLT (\citealt{2006NewAR..50..370S}, \citealt{2010SPIE.7735E..39S}) will introduce a substantial progress in this area. Given the results presented in this paper, increasing by a factor of $\sim$10 the sample of lensing clusters with deep multi-wavelength photometry over a large field of view, similar to our survey in A2667, is expected to set crucial constraints on the intermediate-to-high luminosity domain at z$\sim$7-10, and substantially improve the selection of spectroscopic targets. Such a cluster sample will provide a reference frame allowing to connect the results obtained in blank fields with those obtained in the strongest magnification regimes, as shown in this letter.  

\begin{acknowledgements}
   Part of this work was supported by the French CNRS, the French Programme
National de Cosmologie et Galaxies (PNCG), as well as by the Swiss National Science
Foundation. This work recieved support from the Agence Nationale de la Recherche bearing the
reference ANR-09-BLAN-0234. 
\end{acknowledgements}



\bibliographystyle{aa}  
\bibliography{LFz9} 


\end{document}